\documentclass[12pt, reprint, amsmath,amssymb, aps,prl]{revtex4-1}
\usepackage[dvipdfmx]{graphicx}
\usepackage{ulem}
\usepackage{color}
\usepackage{graphicx}
\usepackage{dcolumn}
\usepackage{bm}
\begin{document}

\title{ Relativistic many body theory of the electric dipole moment of $^{129}$Xe and its implications for probing new physics beyond the Standard Model}

\author{Akitada Sakurai$^{1}$}
\author{B. K. Sahoo$^2$}
\author{K. Asahi$^{1,3}$}
\author{B. P. Das$^1$}
\affiliation{
 $^1$Department of Physics, School of Science, Tokyo Institute of Technology, Ookayama, Meguro-ku, Tokyo 152-8550, Japan\\
 $^2$Atomic, Molecular and Optical Physics Division, Physical Research Laboratory, Navrangpura, Ahmedabad 380009, India	\\
 $^3$RIKEN Nishina Center, 2-1 Hirosawa, Wako-shi, Saitama, 351-0198, Japan
}

\begin{abstract}

We report the results of our theoretical studies of the time-reversal and parity violating electric dipole moment (EDM) of $^{129}$Xe arising from the nuclear Schiff 
moment (NSM) and the electron-nucleus tensor-pseudotensor (T-PT) interaction based on the self-consistent and the normal  relativistic coupled-cluster methods. 
The important many-body effects are highlighted and their contributions are explicitly presented. The uncertainties in the calculations of the correlation and relativistic 
effects  are determined by estimating the contributions of the triples excitations, and the Breit interaction respectively, which together amount to about 0.7\% for the 
NSM and 0.2\% for the T-PT interactions. The results of our present work in combination with improved experimental limits for $^{129}$Xe EDM in the future would 
tighten the constraints  on the hadronic CP violating quantities, and this could provide important insights into new physics beyond the Standard Model of elementary 
particles.
\end{abstract}
\maketitle

The observation of the electric dipole moment (EDM) of a non degenerate system would be a signature of violations of both time-reversal 
({\cal T}) and parity ({\cal P}) symmetries \cite{L.D. Landau, N.F Ramsey}. The CPT theorem implies that T violation amounts to CP 
violation \cite{G. Luders}. The Standard Model (SM) of particle physics contains CP violation in the form of a complex phase in the Kobayashi-Maskawa matrix, which  
however cannot explain the large matter-antimatter asymmetry observed in the Universe \cite{M.B. Gavela}.  This suggests \cite{V. Cirigliano,yoshinaga} that although the SM predicts 
very small values for atomic EDMs, their actual sizes could lie close to the current experimental limits \cite{D.DeMile}. 

The EDMs of diamagnetic atoms have the potential to probe new physics at energy scales much higher than TeV \cite{N.Yamanaka}. They are primarily sensitive to the nuclear 
Schiff Moment (NSM) and the tensor-pseudotensor  (T-PT) electron-nucleus interaction \cite{N.Yamanaka}. The former arises due to CP violating
nucleon-nucleon interactions and the EDMs of nucleons, which at the level of elementary particles arise from CP violating quark-quark 
interactions and the EDMs and chromo-EDMs of quarks \cite{N.Yamanaka}. On the other hand, the latter is due to the T-PT electron-nucleon interaction originating 
from the T-PT electron-quark interaction, which has been predicted by leptoquark models \cite{S.M.Barr}.  

There have been important developments in the search for EDMs of elementary particles and composite systems in recent  years. 
The most stringent  EDM limit to date, $d_\text{Hg} < 7.4\times 10^{-30} e\cdot\text{cm}$ (95\% confidence level (C.L.)), comes from the diamagnetic 
atom, $^{199}$Hg \cite{B. Graner}. This unprecedented precision has been achieved due to the steady improvements 
in the spin precession measurement for this atom over the past three decades. The first result for another diamagnetic atom, $^{225}$Ra,
for which the nuclear octupole deformation is expected to amplify its atomic EDM by about two  - to three orders of magnitude \cite{J.Engle},
was reported 3 years ago \cite{R.H. Parker}  to be $d_\text{Ra} < 1.4\times 10^{-23} e\cdot\text{cm}$ (95\% C.L.). As for the $^{129}$Xe 
diamagnetic atom, three experiments on its EDM are currently under way \cite{W. Heil, T. Sato, F. Kuchler}. Among the above three diamagnetic
species, $^{225}$Ra is radioactive with a half life of 14.9 d, while $^{199}$Hg and $^{129}$Xe are stable. Of the two stable atoms $^{199}$Hg 
and $^{129}$Xe, the latter is characterized by its exceptionally long transverse-spin relaxation times in a gas of atmospheric pressure
\cite{C. Gemmel}. { The result for the  first $^{129}$Xe EDM measurement was published in 1984. \cite{Vold}  } In fact, two groups have reported improved measurements of EDM in $^{129}$Xe recently \cite{Sachdeva,F. Allmendinger}. One 
of these measurements gives its value as $(0.26 \pm 2.33_{\text{stat}} \pm 0.72_{\text{sys}}) \times 10^{-27} e\cdot\text{cm}$ (95\% C.L.)
\cite{Sachdeva}, while the other measurement reports as $(-4.7 \pm 6.4 ) \times 10^{-28} e\cdot\text{cm}$ 
(95\% C.L.) \cite{F. Allmendinger} improving by factors of one-and-half and five times, respectively, than the previous measurement 
$(0.7 \pm 3.3_{\text{stat}} \pm 0.1_{\text{sys}}) \times 10^{-27} e\cdot\text{cm}$ (95\% C.L.) \cite{Rosenberry}.  It is still possible to 
improve its limit by carrying out measurement with a macroscopic number of confined atoms in a glass cell, enabling long-spin coherence times 
and large spin precession signals. The theoretical foundations of Xe EDM were laid in a series of seminal papers by 
Flambaum and co-workers \cite{V.V. Flambaum1, V.V. Flambaum2, V.V. Flambaum3, V.V. Flambaum4}. There have been recent advances in the relativistic many-body calculations of the EDM of this atom\cite{N.Yamanaka,Y.Singh}. The results of these calculations are necessary for extracting
CP violating coupling constants from the measured values of Xe EDM \cite{N.Yamanaka,Y.Singh}. 
The relativistic coupled-cluster (RCC) theory, which is widely considered as the gold standard for the 
electronic structure of heavy atoms \cite{nataraj}, was first applied to $^{129}$Xe EDM by Singh \textit{et al.} by taking 
one particle-one hole (1p1h), two particle-two hole (2p2h) and partial three particle-three hole (3p3h) excitations \cite{Y. Singh3}. 
In the present work, we overcome some of the limitations of the previous calculation by using two different variants 
of the RCC theory. Higher order excitations built from different powers of the 1p1h and 2p2h 
excitations are included in the first approach in a self consistent manner in the evaluation of the EDM, which 
forms a non terminating series. The second approach, which is known as the relativistic normal coupled cluster  
(RNCC) theory, does not treat the bra and ket on the same footing and this enables the expectation value 
representing the EDM to terminate naturally \cite{B.K Sahoo1}.  
We had recently performed ground state electric dipole polarizability calculations of $^{129}$Xe using 
these two methods and obtained results that are in very good agreement with its measured value \cite{A.Sakurai}. 
Given the similarities between the electric dipole polarizability and the EDM from the viewpoint of 
relativistic many-body theory, it is indeed appropriate to apply the two above mentioned RCC methods 
to $^{129}$Xe EDM arising from the NSM and the T-PT electron-nucleus interaction.

The T-PT Hamiltonian is given by \cite{Sandars, V. A. Dzuba, V. V. Flambaum} 
\begin{equation}
	H_\mathrm{int}^\mathrm{TPT} = i  \sqrt{2} G_\mathrm{F} C_\mathrm{T} \sum_{i} (  \bm{\sigma}_\mathrm{N} \cdot {\gamma}_i) \rho_\mathrm{N}(r),
	\label{eq:TPT}
\end{equation}
where $G_\mathrm{F}$ is the Fermi constant, $C_\mathrm{T}$ represents the T-PT coupling constant, $\gamma_i$'s are the Dirac matrices, 
	$\bm{\sigma}_\mathrm{N} = \left( \sigma_x, \sigma_y, \sigma_z\right) $ where $\sigma_x, \sigma_y$ and $\sigma_z$ are
the Pauli spin operators for the nucleus with spin $I = 1/2$, 
and $\rho_\mathrm{N}(r)$ is the nuclear charge density.

The NSM interaction Hamiltonian in the atom is given by \cite{V. A. Dzuba, V. V. Flambaum}
\begin{equation}
	H_\mathrm{int}^\mathrm{NSM} = \frac{3\bm{S} \cdot \bm{r} }{B},
	\label{eq:NSM}
\end{equation}
	where $\bm{S} = S \frac{\bm{I}}{I}$ is the NSM, and $B = \int^{\infty}_{0}dr r^4 \rho_\mathrm{N}(r)$.

In this study, we only consider the first order perturbation in the P and T violating  interaction. 
Therefore, the total atomic Hamiltonian is expressed as
\begin{equation}
	H = H_\mathrm{DC} + \lambda H_\mathrm{PTV}
\end{equation}
where, $H_\mathrm{DC}$ is the Dirac-Coulomb (DC) Hamiltonian that is given by
\begin{equation}
	H_\mathrm{DC} = \sum_{i}^{N_e} [c\bm{\alpha} \cdot \bm{p} + mc^2 \bm{\beta} + V_\mathrm{N} (r_i) ] +\cfrac{1}{2} \sum_{i,j} \cfrac{1}{r_{ij}},
\end{equation}
and $\lambda H_\mathrm{PTV}$ corresponds to either of the P and T violating Hamiltonian given by Eqs. (\ref{eq:TPT}) or (\ref{eq:NSM}).
	Here we assume that the perturbation  parameter $\lambda$ is either $S$ or $G_\mathrm{F} C_\mathrm{T} \langle \bm{\sigma}_\mathrm{N} \rangle$. 
The atomic wave function $| \Psi_0 \rangle$ is written as
\begin{equation}
	| \Psi_0 \rangle \simeq | \Psi_0^{(0)} \rangle + \lambda | \Psi_0^{(1)} \rangle_\lambda,
	\label{eq:psi_0}
\end{equation}	
where the superscripts (0) and (1) represent  the unperturbed and the first order perturbed wave functions due to $H_\mathrm{PTV}$, respectively.

 The expectation value of the EDM in the ground state $|\Psi_0 \rangle $ in an atom in our calculation is given by
 \begin{equation}
 	d_\mathrm{a} = \frac{\langle \Psi_0 | D | \Psi_0 \rangle}{\langle \Psi_0 | \Psi_0 \rangle}
	\label{eq:d_a_H}
 \end{equation}
where, $D$ is the electric dipole moment operator. From Eq. (\ref{eq:psi_0}) and (\ref{eq:d_a_H}), we  can equivalently express 
 \begin{equation}
 	{d_\mathrm{a}} = 2{\lambda} \frac{ \langle \Psi_0^{(0)} |H_\mathrm{PTV}  |\Psi_0^{(1)}\rangle_g }{\langle \Psi_0^{(0)} | \Psi_0^{(0)} \rangle},
	\label{eq:edmdipole}
 \end{equation}
 where $|\Psi_I^{(0)}\rangle_g$ is the first-order perturbed wave function due to the electric dipole, and it is written by
 \begin{eqnarray}
	 | \Psi_0^{(1)} \rangle_g &=&g \sum_{I} |\Psi_I^{(0)}\rangle \frac{\langle \Psi_I^{(0)} | D_g | \Psi_0^{(0)}\rangle}{E_0^{(0)} - E_I^{(0)}}
	\label{eq:psi_1_2}
\end{eqnarray}
with $D_g=D/g$ for an arbitrary parameter $g$. In our calculation, we have used Eq. (\ref{eq:edmdipole}) for the calculation of atomic EDM. 
We present our T-PT and NSM results in terms of $ \eta = \frac{d_\mathrm{a}}{ \langle \sigma_\mathrm{N} \rangle C_\mathrm{T} \times 10^{20} |e| \text{cm}} $ and $\zeta = 
\frac{d_\mathrm{a}}{S\times 10^{17} |e| \text{cm}/(|e|\text{fm}^3)}$ respectively.

The ground state wave function of a closed-shell atom in the RCC theory is expressed as \cite{I. Shavitt}
\begin{equation}
	|\Psi \rangle = e^{T} | \Phi_0 \rangle,
	\label{eq:ccsd}
\end{equation}
where $|\Phi_0\rangle$ is the Dirac-Fock (DF) wave function, the cluster operator $T$ can be written as
 \begin{equation}
 	T = \sum_{I=1}^{N} T_I = \sum_{I=1}^{N} t_I  C_I^{+},
 \end{equation}
 where $I$ is the index for the particle-hole excitation from the DF, closed shell state, $N$ is the maximum value for $I$, $t_I$ is the 
excitation amplitude, and $C_I^+$ is a general $I$ particle-hole excitation operator consisting of a string of creation and annihilation
operators. 
In the singles and doubles approximation in RCC theory (RCCSD method), the maximum value of $I$ 
is restricted to 2; i.e. $T=T_1 + T_2$ , where $T_1$ and $T_2$ are one particle-one hole and two particle-two hole excitation operators. 
We can express $T$ as 
  \begin{equation}
  	T = T^{(0)} + g T^{(1)},
  \end{equation}
 where $T^{(1)}$ is the first-order excitation RCC operator due to $D_g$. Therefore, the total wave function is given by
  \begin{equation}
  	| \Psi_0 \rangle = e^{T^{(0)}+g T^{(1)}} |\Phi_0\rangle.
  \label{eq:ccsd_1}
  \end{equation}
  
 The amplitudes for $T^{(0)}$ can be obtained by solving the equation \cite{R. F. Bishop} 
  \begin{equation}
  	\langle \Phi_0 |C_I^{-}\overline{H}_\mathrm{DC}  |\Phi_0 \rangle = 0,
  \end{equation}
 where $C_I^{-}$, referred to as the de-excitation operators, are the adjoint of  $C_I^{+}$. From here onwards, we use the notation 
$\overline{O} = e^{-T} O e^{T} = (O e^{T})_\text{c}$ for a general operator $O$ and the subscript c stands for the connected terms 
\cite{R. F. Bishop}.  Similarly, amplitudes of  $T^{(1)}$ are obtained by
  \begin{equation}
  	\langle \Phi_0 |C_I^{-} (\overline{H}_\mathrm{DC} T^{(1)} - \overline{D}_g)| \Phi_0\rangle = 0.
  \end{equation}
  
  Using Eqs. (\ref{eq:edmdipole}), (\ref{eq:ccsd}) and (\ref{eq:ccsd_1}), the expression for EDM in the RCCSD method 
can be written as \cite{B.K Sahoo1} 
 \begin{eqnarray}
  \frac{d_\mathrm{a}}{\lambda} &=& 2\langle \Phi_0 |{e^{T^{(0)}}}^{\dagger}H_\mathrm{PTV} e^{T^{(0)}}T^{(1)} | \Phi_0 \rangle_\mathrm{c} \nonumber \\
	 &=& 2 \langle \Phi_0 | [ H_\mathrm{PTV}+( H_\mathrm{PTV} T^{(0)}+ \mathrm{c.c.}) \nonumber \\ && + ({T^{(0)}}^{\dagger} H_\mathrm{PTV} T^{(0)}+\mathrm{c.c.})   \nonumber \\
	 &\quad \quad & + (\frac{1}{2} {T^{(0)}}^{\dagger} H_\mathrm{PTV} {T^{(0)}}^2 + \mathrm{c.c.}) +\cdots ] {T^{(1)}}  |\Phi_0\rangle_\mathrm{c} . \ \ \
	\label{eq:d_a_ccm}
 \end{eqnarray}
 In the relativistic coupled cluster self consistent (RCC(SC)) approach, the combined power of $T^{(0)}$ and its adjoint $T^{(0) \dagger}$ increases in the successive terms in the above expression, and they are computed systematically avoiding double counting till convergence is obtained. In the singles and double approximation, RCC(SC) will be referred to as RCCSD(SC).
 
In order to avoid the non termination problem in the above expression of the RCC method, we use the RNCC method for the evaluation of  
$^{129}$Xe EDM. In this method the RCC bra-state $\langle \Psi | = \langle \Phi_0 | e^{T^{\dagger}}$ is replaced by
  \begin{equation}
  	\langle \widetilde{\Psi} | = \langle \Psi | (1+\widetilde{T}) e^{-T},
	\label{eq:ncc}
  \end{equation}
where $\widetilde{T} = \sum_{I=1}^{N} \widetilde{T}_I = \sum_{I=1}^{N} \widetilde{t}_I C_I^{-}$ is a de-excitation operator with amplitude $ \widetilde{t}_I $, 
similar to $ T^{\dagger} = \sum_{I=1}^{N} T_I^{\dagger} = \sum_{I=1}^{N}t^{*}_I C_I^{-}$. The RNCC bra-state should satisfy
\begin{equation}
	\langle \widetilde{\Psi} | H  = \langle \widetilde{\Psi} | E_0.
	\label{eq:brancc}
\end{equation} 
Furthermore, 
\begin{equation}
	\langle \widetilde{\Psi} | \Psi \rangle = \langle \Phi_0 |(1+\widetilde{T})e^{-T}e^{T}|\Phi_0\rangle = 1 ,
\end{equation}
since the DF state $|\Phi_0 \rangle $ is normalized. Making use of this property, the expectation value of an operator $O$ in the RNCC method can
be expressed as
\begin{eqnarray}
	\langle O \rangle &&= \frac{\langle \widetilde{\Psi} |O| \Psi \rangle }{\langle \widetilde{\Psi} | \Psi \rangle} 
	= \langle \Phi_0(1+\widetilde{T})\overline{O}|\Phi_0\rangle_\mathrm{c}.
\end{eqnarray}
The above expression terminates unlike its counterpart in the RCC.

 In the RNCC method, $\widetilde{T}$ is written as
  \begin{equation}
  	\widetilde{T} = \widetilde{T}^{(0)} + g \widetilde{T}^{(1)},
  \end{equation}
where $\widetilde{T}^{(0)}$  is the  unperturbed de-excitation operator, and  $\widetilde{T}^{(1)}$ is the first order correction to it 
due to $D_g$. Then, the total bra-state can be written as 
\begin{equation}
	\langle \widetilde{\Psi}_0 | = \langle \Phi_0 |(1+\widetilde{T}^{(0)} + g \widetilde{T}^{(1)}) e^{-T^{(0)}-g T^{(1)}} .
\end{equation}
From Eq. (\ref{eq:brancc}), the amplitudes for $\widetilde{T}^{(0)}$ are obtained from
\begin{equation}
	\langle \Phi_0 | (1+\widetilde{T}^{(0)})[ \overline{H}_\mathrm{DC},C_I^+] | \Phi_0 \rangle = 0.
\end{equation}
Similarly, the amplitudes for $\widetilde{T}^{(1)} $  is obtained from
\begin{equation}
	\langle \Phi_0 |[\widetilde{T}^{(1)} \overline{H}_\mathrm{DC} + (1+\widetilde{T}^{(0)}) \{ - \overline{D}_g +(\overline{H}_\mathrm{DC} T^{(1)} )_\mathrm{c} \}  ] C_I^{+} | \Phi_0 \rangle = 0 .
\end{equation}

Adapting Eq. (\ref{eq:d_a_H}) to the case where $D_g$ is a perturbation in the framework of RNCC, we get
\begin{eqnarray}
	\frac{d_\mathrm{a}}{\lambda}  && \equiv \frac{1}{g} \frac{\langle \widetilde{\Psi}_0 |H_\mathrm{PTV}| \Psi_0 \rangle}{\langle \widetilde{\Psi}_0 | \Psi_0 \rangle}  \nonumber \\
		&&= \langle \Phi_0 | \widetilde{T}^{(1)} \overline{H}_\mathrm{PTV} + (1+\widetilde{T}^{(0)})\overline{H}_\mathrm{PTV}T^{(1)} | \Phi_0 \rangle_\mathrm{c} .
\end{eqnarray}
This expression terminates unlike Eq. (\ref{eq:d_a_ccm})  which corresponds to the RCC case. The RNCC method has the merit of satisfying the 
Hellmann-Feynman theorem \cite{R. F. Bishop} in contrast to that of the RCC method.

 In the present study, we have used Gaussian type of orbitals (GTOs) to obtain the DF wave function. The details of the optimized parameters 
that are needed to define the GTOs are discussed in our previous work on the electric dipole polarizability ($\alpha_d$) study on the $^{129}$Xe atom \cite{A.Sakurai}. Using 
these basis functions, we present our results for $\eta$ and $\zeta$ at different levels of approximations of many-body methods in 
Table \ref{tb:result}. One of the methods that has been employed earlier \cite{V. A. Dzuba, Latha}  is the coupled-perturbed Dirac-Fock (CPDF) 
approximation, which takes into account the perturbation of the core to first order in the T and P violating interaction and all orders by the residual 
Coulomb interaction. We had also performed these calculations earlier using the RCCSD method, but considering only some lower order non-linear 
terms in Eq. (\ref{eq:d_a_ccm}) in contrast to the self-consistent procedure in the present work.
The DF contribution as expected is the largest. The CPDF contributions are over  
20\% of the DF results in both the cases. Our DF and CPDF calculations are in good agreement with the 
previous calculations \cite{V. A. Dzuba, Latha}. The correlation effects beyond CPDF, primarily those involving various classes of pair 
correlation collectively reduce the values of $\eta$ and $\zeta$ as reflected in the final results for the two versions of the RCC theory 
used in the present work. We have also given $\alpha_d$ value obtained using our methods in the same table and compared with other available 
results. As can be seen our RCCSD(SC) and RNCCSD calculations for $\alpha_d$ are close to its measured value \cite{Hohm U}. Furthermore,
our RCCSD(SC) results for this quantity as well as $\eta$ and $\zeta$ are in good agreement with those of other calculations of $\alpha_d$
and the two quantities related to EDM, but with different GTO basis functions \cite{Y. Singh3, arXiv:1710.10946v1}. The values of the latter 
two quantities cannot be determined from experiments, but since the two P and T violating interactions related to them have the same rank 
and parity as the electric dipole operator, we expect our calculated values of $\eta$ and $\zeta$ to be accurate. 

 \begin{table}[t]
 \caption{Results for static dipole polarizability $[ e a_0^3 ]$, $\eta$, and $\zeta $ for $^{129}$Xe using different theoretical methods. 
$\Delta_{\text{Breit}}^{\text{CPDF}}$ and $\Delta(T_3)$ are the corrections due to the Breit interaction at the CPDF method and the partial 
triple excitations, respectively.}
\begin{tabular}{llllllll}
\hline \hline
\multicolumn{1}{c}{Method} & \multicolumn{3}{c}{This work}  &  & \multicolumn{3}{c}{Others}  \\ 
\cline{2-4} \cline{6-8} \\
\multicolumn{1}{c}{}           & \multicolumn{1}{c}{$\alpha_d$} & \multicolumn{1}{c}{$\eta$} & \multicolumn{1}{c}{$\zeta$} &  & \multicolumn{1}{c}{$\alpha_d$} & \multicolumn{1}{c}{$\eta$} & \multicolumn{1}{c}{$\zeta$}  \\ \hline
DF 					& 26.87 			& 0.45				& 0.29  	& & 26.87 			& 0.45			& 0.26                        \cite{arXiv:1710.10946v1} \\
					&				& 					&                             		& & 26.92			& 0.45			& 0.29                     	 \cite{Y. Singh3}  \\
CPDF 				& 26.97 			& 0.57 				& 0.38				& & 26.98 			& 0.56			& 0.37                        \cite{arXiv:1710.10946v1} \\
					&				&					&                             		& & 27.7 			& 0.56			& 0.38                     	 \cite{Y. Singh3}  \\
$\Delta^{\text{CPDF}}_{\text{Breit}}$ 			&				&-0.001				& -0.002				&  				& 				&			 		  \\           
RCCSD				& 				& 					&					& & 27.74			& 0.50			& 0.34                     	 \cite{Y. Singh3}  \\                            
RCCSD(SC) 			& 28.12 			& 0.48				& 0.32 				& & 28.13 			& 0.47			& 0.33                       \cite{arXiv:1710.10946v1} \\
$\Delta (T_3)$			& $-0.107$			& $\sim 0$ 	& $\sim 0$	& 				&				&				& 	\\	
RNCCSD 				& 27.51 			& 0.49 				& 0.32				& &  			&  				&                                     \\
Experiment			&   &  			&  				 &    	&   & \multicolumn{2}{l}{27.815(27) \cite{Hohm U}}    \\ \hline \hline
\label{tb:result}
\end{tabular}
\end{table}

 The leading contributions from the terms in Eq. (\ref{eq:d_a_ccm}) are listed in Table \ref{tb:c_cont}. The most important of these is 
$\langle \Phi_0 | H_\mathrm{PTV} T^{(1)}_1 | \Phi_0 \rangle$, which we refer to as the $H_\mathrm{PTV} T^{(1)}_1$ term. It consists of the DF and 
certain classes of correlation effects to all orders in the residual Coulomb interaction such as those represented by the CPDF approximation \cite{A.Sakurai}. 
In particular it subsumes an important correlation effect involving the simultaneous excitation of two core electrons \cite{A.Sakurai}. Its magnitude 
is equal to that of its hermitian conjugate (h.c.).
 The results of our RNCCSD calculations for $\eta$ and $\zeta$  are  given in Table \ref{tb:result}. The breakdown of the contributions from 
the individual terms  are given in Table {\ref{tb:n_cont}}. The leading contributors are $ \langle \Phi_0 |H_\mathrm{PTV} {T_1^{(1)}} | \Phi_0
\rangle $ and $\langle \Phi_0 |\widetilde{T}^{(1)}_1 H_\mathrm{PTV} | \Phi_0 \rangle$. The latter  is the counterpart  of the hermitian
conjugate ($h.c.$) of the former term. The two largest contributions in the case of RCCSD(SC), i.e. $\langle \Phi_0 | H_\mathrm{PTV} T^{(1)}_1 | \Phi_0 \rangle$ and 
$\langle \Phi_0 |  {T^{(0)}_2}^{\dagger} H_\mathrm{PTV} T^{(1)}_1| \Phi_0 \rangle $, and their counterparts for RNCCSD are
not very different. The final results for the two methods given in Table \ref{tb:result} differ by only 2.0\% (T-PT) and are in complete
agreement for the NSM case.
 
\begin{table}
\caption{\,Contributions for T-PT and NSM for $^{129}$Xe  from different terms in RCCSD(SC). }
	\begin{tabular}{lcc}
	\hline \hline
	Leading RCC terms 									&     $\eta$	&   $\zeta$ \\ \hline
	& & \\
	$ H_\mathrm{PTV} {T^{(1)}_1} + \text{h.c.}   $   							&     0.5387	& 	 0.3524\\ 
	$ {T_1^{(0)}}^{\dagger} H_\mathrm{PTV} T^{(1)}_1 + \text{h.c.} $  		&  	 0.0023 	& 	 0.0011\\ 
	$ {T_1^{(0)}}^{\dagger}H_\mathrm{PTV} T_2^{(1)} +\text{h.c.} $  		&  	 $-0.0003$	& 	 0.000036\\ 
	$ {T^{(0)}_2}^{\dagger} H_\mathrm{PTV} T^{(1)}_1+ \text{h.c.} $  		&  	 $-0.0610$ 	& 	$-0.0354$ \\               
	${T_2^{(0)}}^{\dagger}H_\mathrm{PTV} T_2^{(1)} + \text{h.c.} $		&    0.0016	& 	0.000789\\ \hline  \hline   
	\end{tabular}
	\label{tb:c_cont}
	
\caption{\,Contributions for T-PT and NSM for $^{129}$Xe from different terms in RNCCSD.}
\begin{tabular}{lcc}
	\hline \hline
	Leading RNCC terms 							 &   $\eta$		 &  $\zeta$	 \\ \hline 
	 & & \\
	$ H_\mathrm{PTV} {T_1^{(1)}}  $   					&   	0.269	& 	0.176	\\ 
	$ \widetilde{T}^{(1)}_1 H_\mathrm{PTV} $  			&    0.256	& 	0.169	\\ 
	$ \widetilde{T}^{(1)}_1 H_\mathrm{PTV} T^{(0)}_2 $	&  	$-0.029$	& 	-0.017	\\ \hline \hline
	\end{tabular}
	\label{tb:n_cont}
 \end{table}
 
 We have evaluated the numerical error in our RCC calculations by estimating the contributions from the leading electron correlation 
 and relativistic effects that have been omitted in our calculations. The former is characterized by the 3p3h 
 (triples) excitations and the latter by the Breit interaction, which is the leading relativistic correction to the electron-electron 
 Coulomb interaction \cite{IP}. The error due to the 
 first source has been estimated by calculating the perturbed triple excitation amplitudes \cite{A.Sakurai} and the absolute values of 
 this contribution for $\eta$ and $\zeta$ in the present study is $3.9\times 10^{-5}$ and $1.3\times 10^{-4}$, respectively. 
 In this work, the Breit  interaction contributions were found to be $0.6\%$ and $0.9\%$ of the total Dirac Coulomb contributions in the CPDF and RCCSD approximations respectively. Our net
 estimate of the error in our calculations from these two sources are$1.1\times 10^{-3}$ for the T-PT interaction  and $2.1\times 10^{-3}$ 
 for the NSM . 

 The latest reported experimental result for the EDM of $^{129}$Xe is $|d_\mathrm{a}|  <  1.5 \times 10^{-27} |e|$cm 
with 95\% C.L. \cite{F. Allmendinger}. Combining this result with our present RNCCSD calculations,  
$d_\mathrm{a} = 0.49 \times 10^{-20} \langle \sigma \rangle C_\mathrm{T} \ e$cm and $d_\mathrm{a}= 0.32 \times 10^{-17} \ {S/(|e| \ \text{fm}^3)} \ |e|$cm,  
and assuming that the EDM is due to a single source of either the NSM or the T-PT interaction, 
we obtain, respectively, the following upper limits 
\begin{eqnarray}
 |S| < 4.7 \times 10^{-10} \ |e| \text{fm}^3
\end{eqnarray} 
and
\begin{eqnarray}
|C_\mathrm{T}| < 6.1 \times 10^{-7}.
\end{eqnarray}
for the value $\langle \sigma_\mathrm{N} \rangle =1/2$. 

It is important to notice here that the status of nuclear structure calculations for $^{129}$Xe is far more satisfactory than that for 
$^{199}$Hg. The first calculation of the Schiff moment for $^{129}$Xe \cite{Dmitriev} was carried out by taking into full account 
core-polarization effects in the single neutron outside a core approximation of an even-even nucleus. More recently, a substantially improved 
large-scale calculation based on the pair-truncated shell model approach \cite{yoshinaga} has been reported. Results of this calculation are 
of the same sign and of the same order of magnitude as the previous calculation, unlike the case of $^{199}$Hg \cite{N.Yamanaka,J.Engle}. 
Thus, both the atomic and the nuclear calculations are now more reliable for $^{129}$Xe than those for $^{199}$Hg. Turning to the cases of nuclei
exhibiting octupole deformation and vibration collectivities, theoretical calculation of the Schiff moment for $^{129}$Xe could be even more
reliable. The isotopes of such a kind, however, are all found (at least until present) to be unstable, radioactive ones, for which 
experimental precision is largely limited and therefore, reaching the sensitivities to CP violating coefficients of levels achieved by 
$^{199}$Hg and $^{129}$Xe require a long journey of technical developments. Thus, the EDM of $^{129}$Xe atom would be among the leading probes of CP
violating fundamental parameters for the diamagnetic atoms that are under experimental consideration.

It is obvious from the above discussions that the EDM of $^{129}$Xe depends on  two coupling constants $\bar{g}^{(0)}$, $\bar{g}^{(1)}$ [note that the $\bar{g}^{(2)}$ term is expected to be negligible] and one T-PT electron-nucleus coupling constant, $|C_\mathrm{T}|$\cite{Chupp}. 
The predictions for the relative strengths of these three coupling constants vary for different models proposed for new physics.   
Limits for these three coupling constants have been obtained by considering the EDM results for three different systems, one of them being 
$^{129}$Xe \cite{Chupp,Asahi}. Clearly when the sensitivity of $^{129}$Xe EDM experiment improves, the former two limits in tandem with quantum chromodynamics \cite{pospelov1,dekens,vries} will give improved limits for 
$ |\bar{d}_u - \bar{d}_d| $, $|\bar{d}_u + \bar{d}_d|$ and $|\bar{\theta}|$, where $\bar{d}_u$ and $\bar{d}_d$  are the up- and down-quark chromo-EDMs and $\bar{\theta}$ is a parameter associated with CP violation in quantum chromodynamics. 
These new limits are likely to provide useful information on the character of new physics beyond the Standard Model \cite{N.Yamanaka}.
In addition, as mentioned earlier, three experiments on $^{129}$Xe EDM are in progress, and improved limits for this quantity are expected in the near future \cite{T. Sato, F. Kuchler, W. Heil}. These experiments aim to improve the current limit, which is of the order of  $10^{-27} |e|$cm, by as much as three orders of magnitude\cite{F. Kuchler}. If that comes to fruition, then the sensitivity of the $^{129}$Xe EDM experiment could match or even surpass that of the Hg experiment, for which the upper limit ($7.4\times 10^{-30} |e|$cm \cite{B. Graner}) is currently the best  that has been obtained for  any elementary particle or composite systems. It is necessary to emphasize that the theoretical results for $^{129}$Xe EDM reported  in the present work are more accurate and reliable than those obtained for $^{199}$Hg EDM  \cite{sahoo0}. The contributions of the higher order many-body effects are not as large for the former as they are for the
 latter. This is evident from the distinctly smaller differences between the EDM results of the lowest order and self consistent RCCSD levels  for $^{129}$Xe (see Table \ref{tb:result})  compared to those of $^{199}$Hg \cite{sahoo0}. Furthermore, the latter result is in excellent agreement with
 that of the RNCCSD method. Therefore based on both experimental and theoretical considerations, it appears that $^{129}$Xe EDM has the potential to be a more promising candidate for probing new physics beyond the Standard Model than $^{199}$Hg EDM.
 
  The self consistent and the normal versions of the relativistic coupled cluster singles and doubles method have been employed to calculate the ratios of the atomic EDM 
  of $^{129}$Xe to the T-PT interaction coupling constant $\eta$ and the Schiff moment $\zeta$. 
  The results from the two methods disagree only by 2\% for the former and are in perfect agreement for the latter. 
  Comparison of these results with that of the lowest order relativistic coupled cluster singles and doubles method show that the higher order many-body effects converge rapidly, 
  unlike in the case of $^{199}$Hg EDM. For $^{129}$Xe EDM, the estimated errors are 0.2\% for the T-PT interaction and 0.7\% for the NSM. 
  The high accuracy that has been achieved in the present calculations of $\eta$ and $\zeta$
  for $^{129}$Xe suggests that the results of these quantities in combination with the improved results of the future EDM experiments on $^{129}$Xe could serve as a reliable
  probe for new physics beyond the standard model of elementary particle physics.
  
  {
    Computations reported in this work have been performed using Tokyo Institute of the Technology cluster Chiyo and super computer TUBAME 3.0.
    B.K.S. would like to acknowledge use of Vikram-100 HPC of Physical Research Laboratory, Ahmedabad, India for the computations.}

\end{document}